\documentclass[epj]{svjour}
\usepackage{graphics}
\begin{document}
\title{D-mesons in dense nuclear matter}
\author{L.  Tol\'os\inst{1}, J. Schaffner-Bielich\inst{1} \and A. Mishra\inst{2}% etc% \thanks is optional - remove next line if not needed
}                     % Do not remove
\institute{Institut f\"ur Theoretische Physik $\&$ FIAS. J. W. Goethe-Universit\"at. Postfach 11 19  32. 60054 Frankfurt am Main, Germany \and Department of Physics, I.I.T. Delhi, New Delhi - 110 016, India }
\date{Received: date / Revised version: date}
\abstract{ The D-meson properties in dense nuclear matter are studied.
The D-meson spectral density is obtained within the framework of a coupled-channel self-consistent calculation assuming, as bare meson-baryon interaction, a separable potential. The $\Lambda_c(2593)$ resonance is generated dynamically in our coupled-channel model. The medium modifications of the D-meson properties due to Pauli blocking and the dressing of D-mesons, nucleons and pions are also studied. We conclude that the  self-consistent coupled-channel process  reduces the in-medium effects on the D-meson compared to previous literature which do not considered the coupled-channel structure. 
 % end of PACS codes
} %end of abstract
\maketitle
\section{Introduction}
\label{intro}
The study of the properties  of hadrons in hot and dense matter has become a subject of lively interest over the last years in connection with heavy-ion experiments \cite{stateofart} as well as due to implications for astrophysical phenomena. In particular, a strong effort has been invested in understanding the properties of antikaons due to the possibility of kaon condensation in neutron stars \cite{kaplan}. On the other hand, $\rm {K^{\pm}}$ production from nuclear collisions at GSI has shown that in-medium properties of kaons have been seen in different experimental observables like collective flow \cite{lista}.
Medium modifications of $\rm D$ ($\rm {\bar D}$) mesons, which show analogy with $\rm{\bar K}$ (K) coming from the replacement of the $s$ quark ($s$ antiquark) by $c$ quark ($c$ antiquark), have also become a matter of recent interest. The study of medium  modifications of the D-meson could have important consequences for open-charm enhancement in nucleus - nucleus collisions \cite{cassing} as well as for $J/\Psi$ suppression \cite{NA501}.

The NA50 Collaboration \cite{NA50e} has observed an enhancement of dimuons in Pb+Pb collisions 
which has been tentatively attributed to an open-charm enhancement in nucleus - nucleus collisions relative to proton - nucleus reactions at the same $\sqrt{s}$. On the other hand, an appreciable contribution for the $J/\Psi$ suppression is expected to be due to the formation of the quark-gluon plasma \cite{blaiz}. However, the suppression could also be understood in an hadronic environment due to  inelastic comover scattering  in the high-density phase of nucleus - nucleus collisions and, then, the medium modifications of the D-mesons should modify the $J/\Psi$ absorption \cite{blaschke}. 
 Finally, the D-mesic nuclei, which were predicted by the quark-meson coupling  (QMC)  model \cite{qmc}, could also give information about the in-medium properties of the D-meson. It is shown that the $\rm{D^-}$ meson forms narrow bound states with $^{208} Pb$ while the $\rm{D^0}$ is deeply bound in nuclei. Therefore, it is of importance to understand the interaction of the D-meson with the hadronic medium.

At finite density,  the medium modifications have been analyzed via the QCD sum-rule (QSR) approach \cite{arata} as well as using the already mentioned QMC model \cite{qmc}. These models  predict the mass drop of the D-meson to be of the order of 50-60 MeV at nuclear matter density. A similar drop at finite temperature is suggested from the lattice calculations for heavy-quark potentials \cite{digal} together with a recent work based on a chiral model \cite{amruta}. 

In this present paper, the spectral density of a D-meson embedded in dense matter  is shown, incorporating the coupled-channel effects as well as the dressing of intermediate propagators. These medium effects have been ignored in the previous literature. We will show that these effects turn out to  be fundamental for describing the D-meson in dense matter \cite{tolos04}.

\section{The D-meson spectral density}
\label{sec:1}
We present the formalism to obtain the self-energy and, hence, the spectral density of a D-meson embedded in infinite symmetric nuclear matter. For this purpouse, the knowledge of the in-medium DN interaction is required. This amplitude is obtained assuming, as a bare interaction, a separable potential model
\begin{eqnarray}
V_{i,j}(k,k')=\frac{g^2}{\Lambda^2}C_{i,j}\rm{\Theta(\Lambda-k)\Theta(\Lambda-k')} \ ,  
\end{eqnarray}
where $g$ and  $\Lambda$ are the coupling constant and cutoff, respectively. These
two parameters will be determined by fixing the position and the width of the $\Lambda_c(2593)$ resonance. For the interaction matrix $C_{ij}$, we use the result derived from SU(3) flavor symmetry \cite{koch-weise}
. We are, therefore, confronted with a coupled-channel problem since this interaction allows for the transition from DN to other channels, namely, $\pi \Lambda_c$, $\pi \Sigma_c$, $\eta \Lambda_c$ and $\eta \Sigma_c$, all having charm $c=1$. The G-matrix is then given by
\begin{eqnarray}
&&\langle M_1B_1 \mid G(\Omega) \mid M_2B_2 \rangle = \langle M_1B_1
\mid V \mid M_2B_2 \rangle + \nonumber \\
&&\sum_{M_3B_3} \langle M_1B_1 \mid V \mid
M_3B_3 \rangle 
\frac {Q_{M_3B_3}}{\Omega-E_{M_3} -E_{B_3}+i\eta} \times \nonumber\\ 
&&\hspace{3.5cm} \times \langle M_3B_3 \mid
G(\Omega)
\mid M_2B_2 \rangle \ ,
   \label{eq:gmat1}
\end{eqnarray}
with $M_i$ and $B_i$  being the possible
mesons (D, $\pi$, $\eta$) and
baryons ($N$, $\Lambda_c$, $\Sigma_c$), respectively, and their corresponding
quantum numbers, and $\Omega$ is the so-called starting energy. The function $Q_{M_3 B_3}$ stands for the Pauli operator while $E_{M_3 (B_3)}$ is the meson (baryon) single-particle energy (see Ref.~\cite{tolos04} for more details). 

The D-meson single-particle potential is obtained in the Brueckner-Hartree-Fock approach 
\begin{eqnarray}
 &&U_{D}(k,E_{D}^{qp})= \nonumber \\
&&\sum_{N \leq F} \langle D N \mid
 G_{D N\rightarrow
D N} (\Omega = E^{qp}_N+E^{qp}_{D}) \mid D N \rangle,
\label{eq:self}
\end{eqnarray}
where the summation over nucleonic states is limited by the nucleon Fermi momentum. From Eq.~(\ref{eq:self}) one observes that, since the DN  interaction ($G$-matrix) depends on the
D-meson single-particle energy, which in turn depends on the
D-meson potential, we are confronted
with a self-consistent problem. After self-consistency for the on-shell value
$U_{D}(k_{D},E_{D}^{qp})$ is
achieved, one can obtain the full self-energy defined by
\begin{eqnarray}
\Pi_D(k_D,\omega)=2\sqrt{k_D^2+m_D^2} \, U_{D}(k_D,\omega).
\end{eqnarray}
 This self-energy can then be used to
determine the D-meson single-particle propagator
\begin{eqnarray}
&&D_{D}(k_{D},\omega) = \nonumber \\
&&\frac {1}{\omega^2 -k_{D}^2
-m_{D}^2
-2 \sqrt{m_{D}^2+k_D^2} U_{D}(k_{D},\omega)} \ ,
\label{eq:quasip}
\end{eqnarray}
and the corresponding spectral density
\begin{eqnarray}
S_{D}(k_{D},\omega) = -\frac{1}{\pi}{\rm Im\,} D_{D}(k_{D},\omega)
\ .
\label{eq:sqp}
\end{eqnarray}

As  mentioned previously, in our self-consistent scheme, only the value of the potential $U_D$ at the quasiparticle energy has been determined self-consistently. This scheme, in spite of being a simplification, is sufficiently good as already shown in Refs.~\cite{tolos01-02} for the $\rm{\bar K}$ meson.

\begin{figure}
\resizebox{0.45\textwidth}{!}{\includegraphics{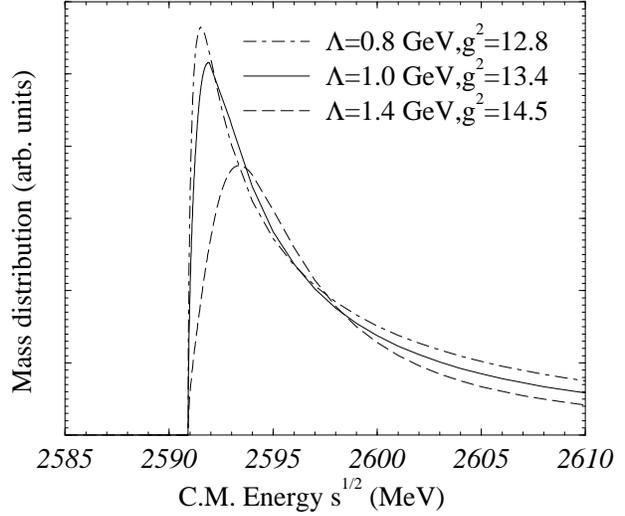}}
\caption{Mass distribution of the $\pi \Sigma$ state as a function of the C.M. energy for a given set of coupling constants and cutoffs, which reflect the $\Lambda_c(2593)$ resonance.}
\label{fig:dmeson1}
\end{figure}

\section{Results}

We start this section by showing in  Fig.~\ref{fig:dmeson1}  the mass distribution  of the $\pi \Sigma_c$ state as a function of the C.M. energy for different sets of  coupling constants $g$ and cutoffs $\Lambda$. This mass distribution is given by 
 \begin{equation}
\frac{d\sigma}{dm} \propto \mid T_{\pi \Sigma_c \rightarrow \pi \Sigma_c}^{I=0} \mid^2
p_{CM} 
\end{equation}
where $p_{CM}$ is the $\pi \Sigma_c$ relative momentum and 
$T_{\pi \Sigma_c \rightarrow \pi \Sigma_c}^{I=0}$ is the isospin zero component of the on-shell s-wave T-matrix for the $\pi \Sigma_c$ channel.
 Our coupled-channel calculation generates dynamically the $\Lambda_c(2593)$ resonance. The position ($2593.9 \pm 2$ MeV) and width ($\Gamma=3.6^{+2.0}_{-1.3}$ MeV) are obtained for a given set of coupling constants and cutoffs running from 0.8 GeV to 1.4 GeV.

\begin{figure}
\resizebox{0.42\textwidth}{!}{\includegraphics{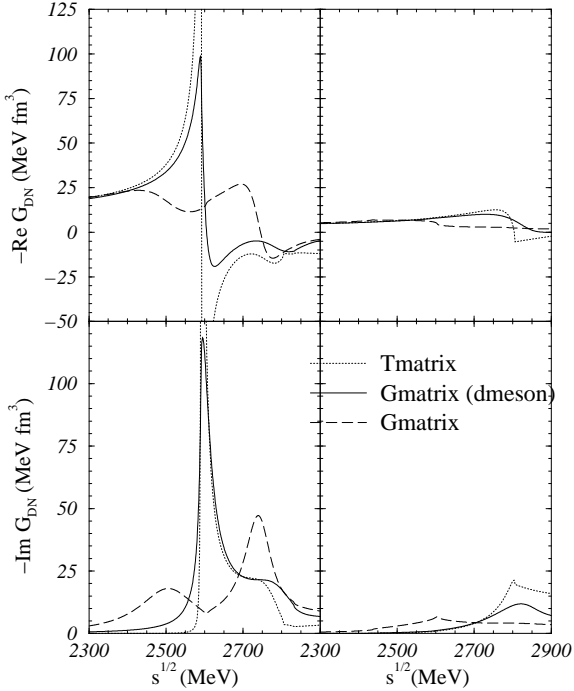}}
\caption{Real and imaginary parts of the DN amplitude for s-wave in the $I=0$ (left panels) and $I=1$ (right panels)  channels as functions of the center-of-mass energy at total momentum zero for $\Lambda=1$ GeV and $g^2=$13.4 and for different approaches: T-matrix calculation (dotted lines), self-consistent calculation for the D-meson at $\rho=\rho_0$ (solid lines), where $\rho_0$ is the nuclear saturation density, and self-consistent calculation for the D-meson including the dressing of nucleons and the pion self-energy at $\rho=\rho_0$ (long-dashed lines). }

\label{fig:dmeson2}
\end{figure}

Once the position and width of the $\Lambda_c(2593)$ resonance are reproduced dynamically, we study the effect of the different medium modifications on the resonance. In  Fig.~\ref{fig:dmeson2} we display  the real and imaginary parts of the s-wave DN amplitude for $I=0$ and $I=1$ within 
different approaches: T-matrix calculation (dotted lines), self-consistent calculation for the D-meson at $\rho=\rho_0$ (solid lines), where $\rho_0$ is the nuclear saturation density, and self-consistent calculation for the D-meson including the dressing of nucleons and the pion self-energy at $\rho=\rho_0$ (long-dashed lines).
When the nucleons and pions are dressed in the self-consistent process, the picture depicted is completely different to the case when only D-mesons are dressed self-consistently. In fact, the DN interaction in $I=0$ becomes smoother in the region of energies where the $\Lambda_c(2593)$ resonance was generated when only the D-mesons were dressed. Furthermore,
we observe one structure  around 2.5 GeV below the $\pi \Sigma_c$ threshold and a second one at 2.8 GeV, which lies below the DN threshold. Both structures are states with the $\Lambda_c$-like quantum numbers.
Whether the first resonant structure  is the in-medium $\Lambda_c(2593)$ resonance 
and the second bump is a new resonance is something that deserve further analysis.

\begin{figure}
\resizebox{0.48\textwidth}{!}{\includegraphics{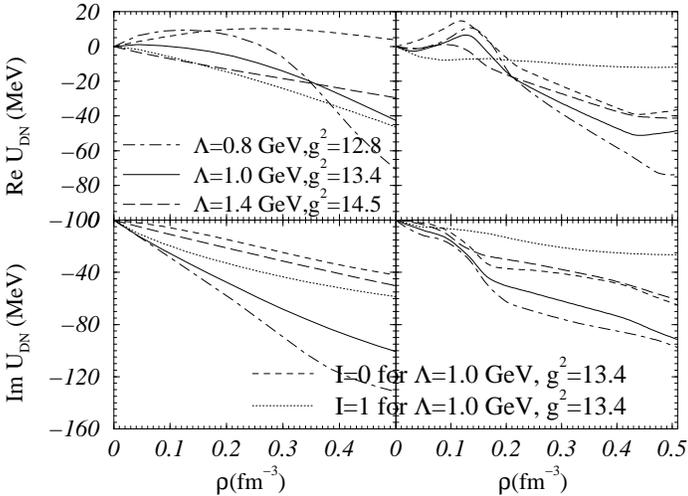}}
\caption{Real and imaginary parts of the D-meson  potential at $k_D=0$ as functions of the density, including the isospin decomposition for $\Lambda=1$ GeV and $g^2=13.4$, for different sets of coupling constants and cutoffs and the two self-consistent approaches discussed in the text: self-consistent calculation for the D-meson (left panels) and self-consistent calculation of the D-meson including the dressing of nucleons and pions (right panels).}
\label{fig:dmeson3}
\end{figure}

The dependence on the cutoff and coupling constants together with the isospin decomposition of the in-medium DN interaction have also  been a matter of study. In Fig.~\ref{fig:dmeson3} we show the  real and imaginary parts of the D-meson potential at $k_D=0$ as functions of the density for the two self-consistent  approaches that were mentioned before: self-consistent calculation for the D-meson (left panels) and self-consistent calculation of the D-meson including the dressing of nucleons and pions (right panels). With regards to the isospin decomposition, when only  D-mesons are dressed, the D-meson potential is governed by the $I=1$ component  while, when nucleons and pions are dressed, the $I=0$ component dominates because of the structure at 2.8 GeV present in the G-matrix. On the other hand, we observe a weak dependence on the chosen set of cutoffs and coupling constants. It is interesting to see that, for any chosen parameters, the coupled-channel effects seem to result in an overall  reduction of the in-medium effects independent of the in-medium properties compared to previous literature \cite{qmc,arata,digal,amruta}. For example, when only the $D$-meson is dressed,  we obtain a range of values for the $D$-meson potential at $\rho=\rho_0$ between  8.6 MeV for $\Lambda=0.8$ GeV and -11.2 MeV for $\Lambda=1.4$ GeV. For the full self-consistent calculation, the range of values covered lies in between 2.6 MeV for $\Lambda=0.8$ GeV  and -12.3  MeV for $\Lambda=1.4$ GeV.

\begin{figure}
\resizebox{0.47\textwidth}{!}{\includegraphics{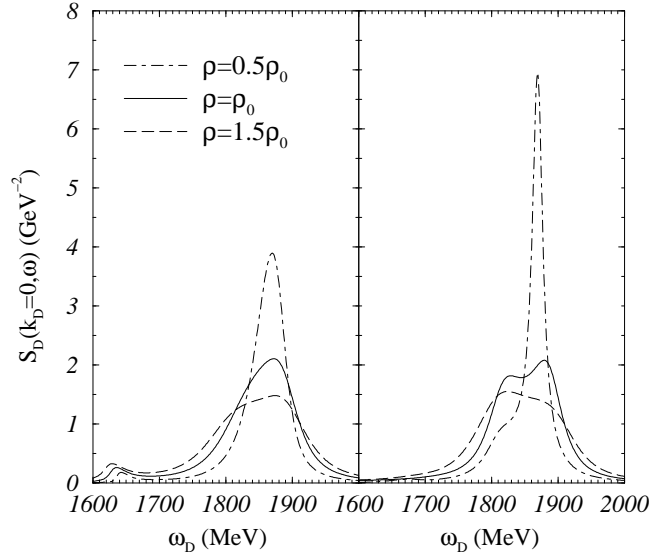}}
\caption{D-meson spectral density at $k_D=0$ as a function of energy with  $\Lambda=1$ GeV and $g^2=$13.4 for different densities and for the two approaches discussed in the text: self-consistent calculation for the D-meson (left panel) and self-consistent calculation of the D-meson including the dressing of nucleons and pions (right panel).}
\label{fig:dmeson4}
\end{figure}

Finally, once the self-consistency is reached, we calculate the full D-meson self-energy and the corresponding spectral density. The spectral density at zero momentum is shown in Fig.~\ref{fig:dmeson4} for $\Lambda=1$ GeV and $g^2=13.4$, and for several densities in the  two self-consistent approaches considered before. When only the D-meson is dressed self-consistently (left panel of Fig.~\ref{fig:dmeson4}), the quasiparticle peak moves slightly to  lower energies as density increases since the D-meson potential becomes more attractive (see left panels of Fig.~\ref{fig:dmeson3}). The $\Lambda_c$ resonance is seen for energies around 1.63-1.65 GeV as a second peak on the left-hand side of the quasiparticle peak. For the second approach when nucleons and pions are dressed (right panel of Fig.~\ref{fig:dmeson4}), the structure around 2.8 GeV of Fig.~\ref{fig:dmeson1} mixes with the quasiparticle peak which translates into a broadening of the spectral density at the quasiparticle energy.

\section{Concluding remarks}

We have performed a microscopic self-consistent coupled-channel calculation of the single-particle potential and, hence, the spectral density of a D-meson embedded in symmetric nuclear matter taking, as bare interaction, a separable potential \cite{tolos04}. The $\Lambda(2593)_c$ resonance has been obtained dynamically. 
We have also studied the medium effects on that resonance and, therefore, on the D-meson potential coming from the Pauli blocking and the dressing of  nucleons and pions.  We have concluded that, independently of the medium properties of the intermediate states, the self-consistent coupled-channel effects result  in a small attractive real part of the in-medium D-meson potential. However, the production of D-mesons in the nuclear medium
will be still enhanced due to the broad D-meson
spectral density.  This effect is similar to the one obtained for the enhanced $\rm \bar K$ production in heavy-ion collisions
\cite{Toloseffect}, where the overlap of the Boltzmann factor with the strength in the low-energy region of the $\rm \bar K$  spectral density turned out to be a source of additional attraction and, hence, increased the produced number of $\rm \bar K$.  The in-medium effects seen in this work can be studied with the future PANDA experiment at GSI \cite{ritman}. In this experiment, in-medium changes of the open charm hadrons will be addressed by the study of the excitation function and the correlation function of $\rm D$ and $\rm{\bar D}$ mesons.

\section*{Acknowledgment}

L.T. wishes to acknowledge financial support from the Alexander von Humboldt Foundation.

\end{document}